\documentclass[epj]{svjour}

\usepackage{latexsym}
\usepackage{graphics}
\usepackage{epsfig}

\usepackage{dcolumn}
\usepackage{bm}

\begin{document}

\title{The Abundance of New Kind of Dark Matter Structures}

\author{Yupeng Yang\inst{1,2} \and Xuelei Chen\inst{3} \and Tan Lu\inst{1,2,4} 
\and Hongshi Zong\inst{1,2} 
\institute{Department of Physics, Nanjing University, Nanjing, 210093, China 
\and Joint Center for Particle, Nuclear Physics and Cosmology, Nanjing, 210093, China
\email{yyp@chenwang.nju.edu.cn}
\and National Astronomical Observatories, Chinese Academy of Sciences, Beijing, 100012, China
\and Purple Mountain Observatory, Chinese Academy of Sciences, Nanjing, 210008, China}}

\date{Received: date / Revised version: date}

\abstract{A new kind of dark matter structures, ultracompact minihalos (UCMHs) 
was proposed recently. They would be formed during the radiation 
dominated epoch if the large density perturbations are existent. 
Moreover, if the dark matter is made up of 
weakly interacting massive particles, the UCMHs 
can have effect on cosmological evolution 
because of the high density and dark matter annihilation within them. 
In this paper, one new parameter is introduced to consider
the contributions of UCMHs due to the dark matter annihilation 
to the evolution of cosmology, 
and we use the current and future CMB observations 
to obtain the constraint on the new parameter 
and then the abundance of UCMHs.
The final results are applicable for a wider range 
of dark matter parameters.}

\PACS{{PACS-key}{discribing text of that key}   \and
      {PACS-key}{discribing text of that key}} 
 
\maketitle

\section{Introduction}
\label{intro}
It is well known that the present structures of Universe 
originate from the density 
perturbations ($\delta\rho/\rho \sim 10^{-5}$). 
Theoretically, primordial black halos (PBHs) 
can be formed if the density perturbations 
are larger than 0.3 \cite{PHBs}. 
Recently, it was proposed that a new kind of dark matter structures, 
ultracompact minihalos (UCMHs), can
be formed during earlier epoch if 
the density perturbations are between 
3$\times 10^{-4}$ and 0.3 \cite{ucmhs}. 
These large density perturbations can be obtained 
through the phase transitions in the early Universe \cite{schmid}.

Although the presence of dark matter has
been shown by many observations, its nature
still remains unknown. 
At present there are many dark matter models and 
the weakly interacting massive particles (WIMPs) 
model is a frequently studied one \cite{wimp_1,wimp_2,wimp_3}. 
These dark matter particles can annihilate 
into standard particles such as electrons,
positrons, protons, antiprotons, or photons. 
Therefore, the evolution of the cosmos can be influenced 
by the dark matter annihilation process due to 
the interaction between the particles \cite{zl}. 
On the other hand, since the dark matter 
annihilation rate is proportional to the 
square of the number density, the UCMHs 
would have effects on the cosmological evolution 
and can be regarded as the $\gamma$-ray sources 
due to the higher density within them. 
This might be a way for 
the indirect search of dark matter \cite{9712318,0207125,0806.2911}. 
In Ref. \cite{scott}, the authors calculated 
the $\gamma$-ray flux from the nearby UCMHs 
which have been formed during three phase transitions: 
electroweak symmetry breaking (EW), QCD confinement (QCD), 
and $e^{+}e^{-}$ annihilation ($e^{+}e^{-}$). 
They found that after considering the sensitivity of EGRET or Fermi,
 these objects should be detectable. 
In Ref. \cite{josan} the authors investigated the constraint 
on the current abundance of UCMHs. 
They found that the tightest bound is 
$f_{UCMHs} \leq 10^{-5}$ 
for $M_{UCMHs} \sim 10^{5} M_{\odot}$ 
if no gamma-ray emission is detected from UCMHs.

In Ref. \cite{yyp}, the authors obtained 
the current abundance of UCMHs using the WMAP-7 years 
data, where the contributions from halos and subhalos 
are also included. Therefore, in addition to 
the parameter which describes the current abundance 
of UCMHs, another one which considers the nature of
dark matter must be added. 
In this paper, we only consider the UCMHs effect simply. 
Its advantage is that besides the cosmological parameters
only one free parameter is needed. 
Although we do not get the accurate results 
due to the rejection of the annihilation effect from 
the classical dark matter halos, we found that the comparable results 
can also be obtained, and they can be applied for 
wider range of dark matter parameters. 
In order to get the constraint on the abundance of UCMHs 
we use the WMAP-7 years data. Moreover, we also 
want to discuss the constraint for the future 
CMB observation of Planck. In this paper, 
we use the WMAP-7 years results as the fiducial model and produce 
the mock data for the future Planck-3 years observations, 
and then use these data to get the results.

This paper is organized as follows. In Sec. II we give the relevant equations
which describe the UCMHs and their contributions 
to the cosmological evolution. 
In Sec. III, we give our results of the constraint on
the current abundance of UCMHs using the current CMB
data and the future 3-years observation of Planck.
The conclusion is presented in Sec. IV.

\section{The constraint on the abundance of UCMHs from 
the current and future CMB observations}
\subsection{The relevant equations}

After the formation of UCMHs, 
dark matter particles can be accreted by radial infall 
and the mass of UCMHs evolves according to \cite{scott}
\begin{equation}
\label{Mh}
M_\mathrm{UCMHs}(z) = \delta m \left(\frac{1 + z_\mathrm{eq}}{1+z}\right),
\end{equation}
where $\delta m$ is the mass contained within a perturbation 
at the redshift of matter-radiation equality $z_\mathrm{eq}$. 
As in Ref. \cite{scott} (also followed by Ref. \cite{yyp}), in this paper we adopt 
$\delta m$ = [$5.6 \times 10^{-19}, 
1.1 \times 10^{-9} $, 0.33]$M_\odot$
for the three phase transitions: 
electroweak symmetry breaking,
QCD confinement and $e^{+}e^{-}$ annihilation.
 
The density profile of UCMHs is \cite{scott}
\begin{equation}
\label{density}
\rho_{UCMHs}(r,z) = \frac{3f_\chi M_\mathrm{UCMHs}(z)}{16\pi R_\mathrm{UCMHs}(z)^\frac{3}{4}r^\frac{9}{4}},
\end{equation}
where ${R_\mathrm{NACHOs}(z)} = 
0.019\left(\frac{1000}{z+1}\right)\left(\frac{M_\mathrm{UCMHs}(z)}
{\mathrm{M}_\odot}\right)^\frac{1}{3} \mathrm{pc}$ 
and $f_{\chi}$ is the dark matter fraction.
We follow the assumption that UCMHs 
stop growing at $z \approx 10$ because 
the structure formation process prevents further accretion 
after the redshift. 

Similar to the case of the black hole \cite{0912.5297}, 
and following Ref. \cite{yyp}, we assume that 
the UCMHs have a monochromatic mass function, 
and the abundance of UCMHs is the same 
everywhere and they do not merger with others 
\footnote{In fact, the abundance of UCMHs 
can be different in different places. 
The higher density perturbations from 
which UCMHs form will be highly clustered, 
and in this case the number 
of UCMHs should be treated as the average one.}.
We neglect the energy loss of the dark matter 
annihilation production within UCMHs \cite{1003.3466}. 
We also neglect 
the effect of adiabatic contraction 
on the density profile of UCMHs. 
This is 
because it has remarkable effect 
only around the edge of baryonic core, 
and the density profile of the centre 
is not changed significantly \cite{scott}.
Following Ref. \cite{yyp}, based on these assumptions, 
we can get the annihilation rate of UCMHs
\begin{eqnarray}
\label{eq:fai}
\Gamma
= \frac{f_{UCMHs} \rho_{0,critical}}{M_{UCMHs}(z=0)}(1+z)^3 
\frac{\langle\sigma v\rangle}{m_\chi^2} \int 4\pi r^2\rho^2(r,z)\mathrm{d}r,
\end{eqnarray}
where $\Gamma$ is the annihilation rate per unit volume of UCMHs. 
$N_{UCMHs}$ is the number density of UCMHs, 
$f_{UCMHs}$ = $\rho_{0,UCMHs}/\rho_{0,critical}$ is the current 
abundance of UCMHs.   
The upper limit of the integral is $R_{UCMHs}$. 
However, there is a cut radius
$r_{cut}$ due to the dark matter annihilation. $r_{cut}$ can be estimated as follows.
According to Ref. \cite{rcut}, 
the evolution equation of the dark matter number density can be written as 
\begin{equation}
\frac{d{n}_\chi(r,t)}{dt}=-\langle\sigma v\rangle n^2_\chi(r,t),
\end{equation} 
then at any time $t$, the number density is
\begin{equation}
n_\chi(r,t)=\frac{n_\chi(r,t_i)}
{1+n_{\chi}(r,t_{i})\langle\sigma v\rangle(t-t_i)}.
\end{equation} 

 we can get the maximal density 
$\sim m_{\chi}/\langle\sigma v\rangle(t-t_{i})$.  
We define $r_{cut}$ at present time $t_{0}$ which 
satisfies the equation 
\begin{eqnarray}
\label{eq:r_cut}
\rho(r_{cut}) = \frac{m_{\chi}}{\langle \sigma v \rangle(t_{0}-t_{i})},
\end{eqnarray}
where $t_{0} \approx 13.7 Gyr$ \cite{scott,josan} 
is the age of the universe, 
$t_{i}$ is the time of UCMHs formation and as in Ref. \cite{josan}, we choose 
$t_{i}(z_{eq}) \approx 77kyr$.
Following \cite{scott,josan}, 
we assume that the density is constant within 
$r_{cut}$, $\rho(r \le r_{cut}) = \rho(r_{cut})$.

\par

Considering the dark matter annihilation, the evolution 
of ionization fraction $x_{e}$ is \cite{zl}
\begin{equation}
(1+z){dx_e\over dz} = {1\over H(z)}[R_s(z)-I_s(z)-I_\chi(z)],
\end{equation}
where $R_{s}$ is the standard recombination rate, which is the
ionization rate by standard sources, and $I_{\chi}$ is the ionization
rate sourced by dark matter. $I_{\chi}$ is given as \cite{zl}
\begin{equation}
{I_\chi} = {\chi_i f_{\chi}{2 m_\chi c^{2}\over n_b E_b}}{\Gamma},
\end{equation}
where $n_b$ is the baryon number density, $E_b = 13.6 eV$ 
is the ionization energy, $m_{\chi}$ is the dark matter mass, and 
$f_{\chi}$ is the released energy fraction 
deposited in the baryonic gas
during the annihilation. Following Ref. \cite{yyp}, 
we set $f_{\chi} = 1$. 
$\chi_i$ is the energy fraction which 
ionizes the baryonic gas and we  
accept the form given in Ref. \cite{cxl}, 
followed by Refs. \cite{zl,yyp} 
\begin{equation}
{\chi_i} = {\left(1 - {x_e}\right) / 3},
\end{equation}  
where $x_e$ is the fraction of free electrons.

In this paper, we introduce one new parameter $f_{\star}$
which describes the nature of dark matter 
and the current abundance of UCMHs 
\begin{equation}
f_{\star} = \left(\frac{\langle\sigma v\rangle}{10^{-26}cm^{3}s^{-1}}\right)
\left(\frac{m_{\chi}}{1GeV}\right)^{-1} \left(\frac{f_{UCMHs}}{10^{-5}}\right)
\end{equation} 

This is different from Ref. \cite{yyp} where $f_{UCMHs}$ 
and one of $\langle\sigma v\rangle$ and ${m_{\chi}}$  
must be treated as free parameters  
in order to include the contributions of halos. 
For example, they fix the value of $\langle\sigma v\rangle$ 
and take the dark matter mass and current abundance of UCMHs 
as free parameters. We will see that the neglect 
of contributions from halos do not affect the final results significantly, 
and they can be used for wider range of dark matter parameters.

\subsection{Constraint on the abundance of UCMHs}

We have modified the public 
COSMOMC code \cite{cosmomc} in order to 
include the new parameter $f_{\star}$. 
We use the WMAP-7 years data and the mock data of future 3-years 
observations of Planck \cite{planck} to get 
the constraint on the abundance of UCMHs. 
In order to produce the mock data, 
we use the WMAP-7 years results \cite{wmap7} as the fiducial model. 
We consider 6 cosmological parameters and the new parameter: 
\begin{equation}
\{\Omega_b h^2,\Omega_d h^2,\theta,\tau,n_s,A_s,f_{\star}\},
\end{equation}
where $\Omega_b h^2$ and $\Omega_d h^2$ are the density of 
baryon and the density of dark matter, respectively, $\theta$ is the ratio of
the sound horizon at recombination to its angular diameter distance 
multiplied by 100, $\tau$ is the optical depth, and
$n_s$ and $A_s$ are the spectral index and amplitude of
the primordial density perturbation power spectrum, respectively.
Because the annihilation rate of UCMHs are almost the same 
for the three phase transitions, we only do the calculation for 
the $e^{+}e^{-}$ case and the results are listed in Table 1. 

\begin{table}
\caption{$2\sigma$ posterior constraint on 
$f_{\star}$ and the related 
cosmological parameters,  $\tau$ (the optical depth)
and $z_{re}$ (the redshift of reionization) 
are listed for the $e^{+}e^{-}$ case. 
We used the WMAP-7 years data and the mock data (MD) for the future 3-years 
observation of Planck.}

\label{tab:cons}       
\begin{center}
\begin{tabular}{llll}
\hline\noalign{\smallskip}
Parameter & $f_{\star}$ & $\tau$ & $z_{re}$ \\
\noalign{\smallskip}\hline\noalign{\smallskip}
WMAP-7     & 18.5 & 0.098 & 11.9 \\
Planck-3(MD)  & 1.9 & 0.086 & 9.75  \\
\noalign{\smallskip}\hline
\end{tabular}
\end{center}
\vspace*{0.3cm}  
\end{table}
%
%
In Tab.~\ref{tab:cons}, the $2\sigma$ value 
of the parameters are shown.  
>From this it can be seen that for the future 3-years observation of Planck 
the constraint is   
better than the WMAP-7 years data by about one order of magnitude. 

For the WMAP-7 years data, the constraint 
on the current abundance of UCMHs is
\begin{equation}
f \leq 1.85 \times 10^{-4} \left(\frac{\langle\sigma v\rangle}{10^{-26}cm^{3}s^{-1}}\right)^{-1}
\left(\frac{m_{\chi}}{1GeV}\right),
\end{equation}
while for the future 3-years observations of Planck (mock data),  
the constraint is 
\begin{equation}
f \leq 0.19 \times 10^{-4} \left(\frac{\langle\sigma v\rangle}{10^{-26}cm^{3}s^{-1}}\right)^{-1}
\left(\frac{m_{\chi}}{1GeV}\right).
\end{equation}

These constraints are comparable with the results in Ref. \cite{yyp}. 
For this point, we think that the contributions 
of the conventional dark matter halos are smaller 
than those of the UCMHs, and the dominant effect is from UCMHs. 
This is also shown in Ref. \cite{zhang} where the luminosity 
from UCMHs is dominant compared with the background.  
Moreover, we do not have to set 
the range of dark matter mass for 
the new parameter introduced by us, 
so the constraints in this paper are applicable 
for wider range of dark matter mass. 

\section {Conclusion} 
During an earlier epoch, 
if the density perturbations are between 3$\times 10^{-4}$ and 0.3,  
a new kind of dark matter 
structure, ultracompact minihalos 
would be formed. If the dark matter is made up of weakly interacting massive particles, 
these objects would have effects 
on the cosmological evolution, especially 
the recombination and reionization. 
Therefore, the CMB observations can give 
constraints on their abundance. 
In this paper, 
we reinvestigate the current abundance 
of them using the WMAP-7 years data and
the mock date for future 3-years observation of Planck. 
Our treatment is different form Ref. \cite{yyp},
Although we do not include the halos' contributions,
we still find that the results are comparable.
It is because the contributions from UCMHs 
to the evolution of cosmology are dominant.
Moreover, because only one new parameter 
is introduced, so the final results 
are applicable for wider range of 
dark matter mass. 
\par

From these results we can see that for the typical value 
$\langle\sigma v\rangle = 3 \times 10^{-26} cm^{3}s^{-1}$ 
and $m_{\chi} = 100 GeV$, the current abundance of UCMHs 
are $f \leq 6.2 \times 10^{-3}$ and $6.3 \times 10^{-4}$ for 
the WMAP and Planck, respectively. 
On the other hand, because the parameter 
$\langle\sigma v\rangle$ is not fixed in this paper, 
so we can get the constraint on $f_{UCMHs}$ 
for the abnormal value of $\langle\sigma v\rangle$.
For example, the recent observations of the cosmic ray, 
such as PAMELA \cite{pamela} and ATIC \cite{atic}, 
$\langle\sigma v\rangle \sim 10^{-23} cm^{3}s^{-1}$
and $m_{\chi} \sim 1 TeV$, the current abundance of UCMHs
are $f \leq 1.85 \times 10^{-4}$ and $1.9 \times 10^{-5}$ 
for the WMAP and Planck, respectively.

\section{Acknowledgments}
We thank Mingzhe Li and Weimin Sun 
for helpful suggestions and improving the manuscript.
Our MCMC chains computation was performed on the Shenteng 7000 system of the Supercomputing 
Center of the Chinese Academy of Sciences. This work is supported in part 
by the National Natural Science Foundation of China (under Grant Nos 10775069, 10935001 and
11075075) and the Research Fund for the Doctoral Program of Higher
Education (under Grant No 200802840009)

\newcommand\PRD[3]{~Phys. Rev. D{\bf ~#1}, (#2)~#3}
\newcommand\APJ[3]{~Astrophys. J.{\bf ~#1}, (#2)~#3}
\newcommand\PRL[3]{~Phys. Rev. Lett.{\bf ~#1}, (#2)~#3}
\newcommand\PR[3]{~Phys. Rept.{\bf ~#1}, (#2)~#3}
\newcommand\NJP[3]{~New J.Phys.{\bf ~#1}, (#2)~#3}
\newcommand\APJL[3]{~Astrophys. J. Lett.{\bf ~#1}, (#2)~#3}
\newcommand\Nature[3]{~Nature{\bf ~#1}, (#2)~#3}
\newcommand\AP[3]{~Astropart. Phys{\bf ~#1}, (#2)~#3}
\newcommand\JCAP[3]{~JCAP{\bf ~#1}, (#2)~#3}
\newcommand\APJS[3]{~Astrophys. J. Suppl.{\bf ~#1}, (#2)~#3}
\newcommand\MNRAS[3]{~MNRAS{\bf ~#1}, (#2)~#3}

\end{document}